\documentclass[%
 reprint,
superscriptaddress,
 amsmath,amssymb,
 aps,
pra,
]{revtex4-2}

\usepackage{graphicx}
\usepackage{dcolumn}
\usepackage{bm}
\usepackage{hyperref}
\usepackage{here}

\usepackage{color}

\begin{document}

\preprint{APS/123-QED}

\title{Optical trapping of the transversal motion for an optically levitated mirror}

\author{Takuya~Kawasaki}
  \email{takuya.kawasaki@phys.s.u-tokyo.ac.jp}
 \affiliation{Department of Physics, University of Tokyo, Bunkyo, Tokyo 113-0033, Japan}
\author{Naoki~Kita}%
 \affiliation{Department of Physics, University of Tokyo, Bunkyo, Tokyo 113-0033, Japan}
\author{Koji~Nagano}
 \affiliation{Institute of Space and Astronautical Science, Japan Aerospace Exploration Agency, Sagamihara, Kanagawa 252-5210, Japan}
\author{Shotaro~Wada}
 \affiliation{Department of Physics, University of Tokyo, Bunkyo, Tokyo 113-0033, Japan}
\author{Yuya~Kuwahara}
 \affiliation{Department of Physics, University of Tokyo, Bunkyo, Tokyo 113-0033, Japan}
\author{Masaki~Ando}
 \affiliation{Department of Physics, University of Tokyo, Bunkyo, Tokyo 113-0033, Japan}
 \affiliation{Research Center for the Early Universe (RESCEU), University of Tokyo, Bunkyo, Tokyo 113-0033, Japan}
\author{Yuta~Michimura}
  \email{michimura@phys.s.u-tokyo.ac.jp}
 \affiliation{Department of Physics, University of Tokyo, Bunkyo, Tokyo 113-0033, Japan}

\date{\today}

\begin{abstract}
Optomechanical systems are suitable for elucidating quantum phenomena at the macroscopic scale in the sense of the mass scale.
The systems should be well-isolated from the environment to avoid classical noises, which conceal quantum signals.
Optical levitation is a promising way to isolate optomechanical systems from the environment.
To realize optical levitation, all degrees of freedom need to be trapped.
Until now, longitudinal trapping and rotational trapping of a mirror with optical radiation pressure have been studied in detail and validated with various experiments.
However, less attention has been paid to
the transversal trapping of a mirror.
Herein, we report a pioneering result where we experimentally confirmed
transversal trapping of a mirror of a Fabry-P\'erot cavity using a torsional pendulum.
Through this demonstration, we experimentally proved that optical levitation is realizable with only two Fabry-P\'erot cavities that are aligned vertically.
This work paves the way toward optical levitation and realizing a macroscopic quantum system.

\end{abstract}

\maketitle


\section{\label{sec:level1}Introduction}
Quantum mechanics is an established theory in modern physics.
However, no experiment could observe the quantum phenomena of a massive object over the Planck mass, and it is not entirely clear what makes classical systems classical.
Thus, there are active discussions about whether quantum mechanics breaks at some mass scale.
One possible scenario is that quantum mechanics is valid over all mass scales, but macroscopic systems tend to couple strongly to the environment.
The interaction between the macroscopic systems and the environment causes thermal decoherence.
Because of this thermal decoherence, the systems lose their quantumness.
Other scenarios involve the need to modify the quantum mechanics or inclusion of gravitational effects.
For example, the continuous spontaneous localization model~\cite{Bassi:2012bg} is a theory that modifies quantum mechanics by adding a collapse mechanism.
Gravitational decoherence is also one of the decoherence models~\cite{Diosi:1988uy, Penrose:1996cv}.

These discussions lead to a demand for the realization of a macroscopic quantum system to test such proposed models.
In addition to various decoherence mechanisms, the effect of gravity is crucial in a macroscopic quantum system.
Thus, a macroscopic quantum system can be a platform to experimentally elucidate quantum gravity and ultimately lead to the unification of the theories of gravity and quantum mechanics~\cite{Marletto:2017kzi, Bose:2017nin, Belenchia:2018szb, Carney:2019, Christodoulou:2018cmk, Vedral:2020dnh}.

Recent progress in experimental techniques provides a suitable opportunity to realize a macroscopic quantum system.
In particular, optomechanical systems have been increasingly attracting attention as promising candidates for macroscopic quantum systems.
In an optomechanical system, a mechanical oscillator is coupled to an optical field, and optical interferometry enables precise measurement of the displacement of the oscillator.
The sensitivity is high enough to reach the quantum regime.
With an optomechanical system in the quantum regime, it will be possible to test modified theories of quantum mechanics by using the superposition state of the mechanical oscillator.

Until now, below nanogram scales, several systems have been reported to reach the quantum regime~\cite{chan2011laser, teufel2011sideband, peterson2016laser} in the sense that they attain the standard quantum limit.
Though there have been experimental studies with even heavier mechanical oscillators in the range of micrograms to kilograms, no such system is reported to have reached the quantum regimes~\cite{Pontin:2018uud, Matsumoto:2018via, Komori:2019zlg, MowLowry:2008zz, Corbitt:2007, Abbott:2009zz, Yu:2020ece, Michimura:2020yvn}.
As the oscillator mass increases, thermal decoherence caused by the mechanical structure that supports the oscillator prevents the system from reaching the quantum regime.
In other words, thermal noise conceals the detection of quantum signals in massive oscillators.

\begin{figure}
    \centering
        \includegraphics[width=0.8\columnwidth,clip]{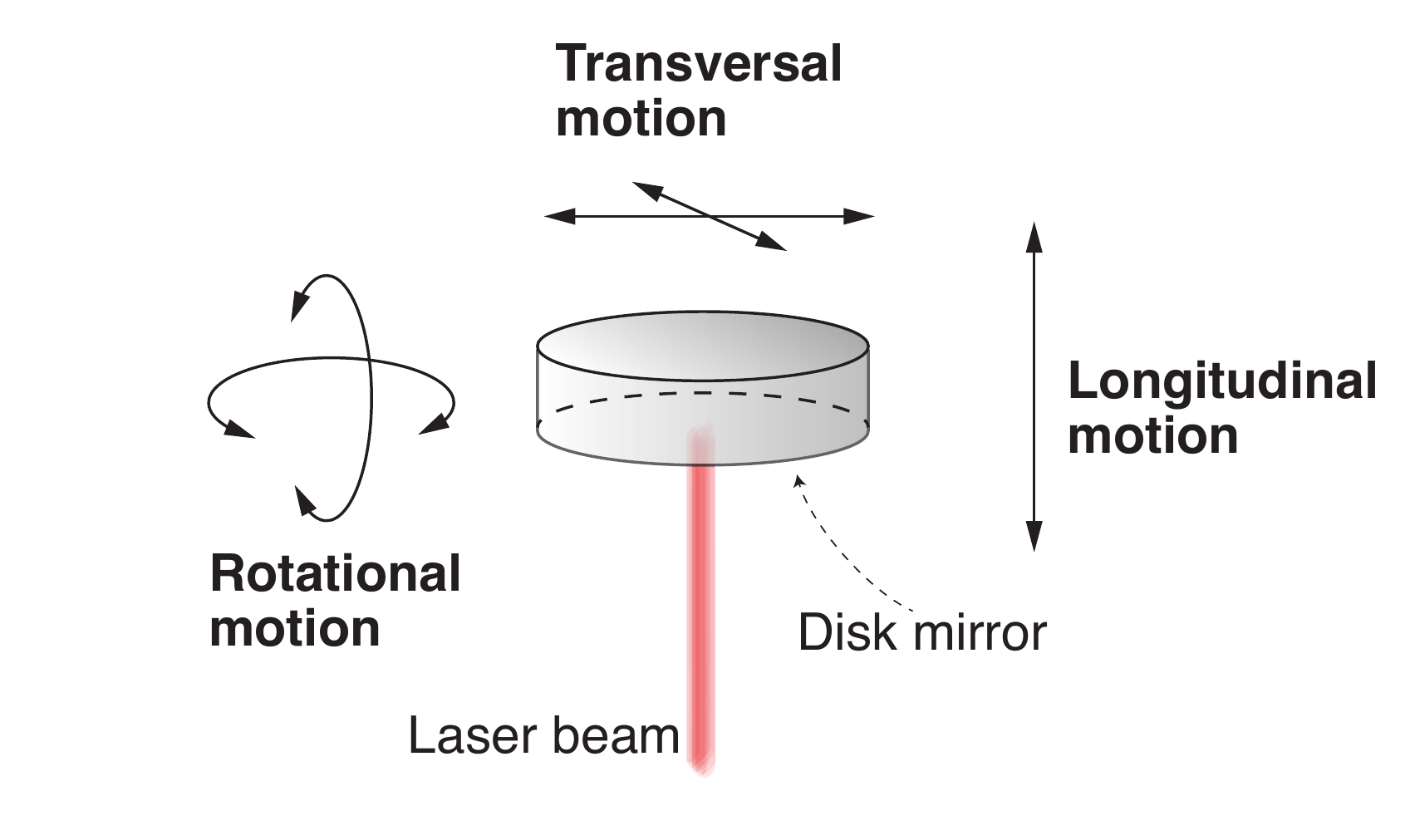}
        \caption{Degrees of freedom used in this study.
        The longitudinal direction is the direction in which we measure the displacement of the oscillator with laser beams.
        The transversal direction is defined to be transverse to the longitudinal direction.}
        \label{fig:direction}
\end{figure}
Among the efforts in achieving high mechanical quality factor to lower the 
thermal noise, optical levitation is a suitable alternative to eliminate thermal noise from mechanical support.
The thermal noise from such mechanical structures can be avoided by supporting the mirror only by the radiation pressure of the laser lights.

A widely used method to levitate an object employs optical tweezers~\cite{Ashkin:1971, Grier:2003, Chang:2010, Li:2013, Tebbenjohanns:2020yek, Delic:2020}.
Optical tweezers use the optical gradient force.
Thus, the objects need to be inside a highly focused laser beam, to be levitated.
This limits the mass of the objects that can be levitated using optical tweezers to nanogram-scale.
An alternative way to levitate more massive objects is to levitate a highly reflective mirror by the radiation pressure of laser lights~\cite{Singh2010, guccione, Michimura:2017zfe}.

For any type of optical levitation, stable trapping of all degrees of freedom of the mirror is necessary.
Here, the direction in which we measure the displacement of the oscillator is called the longitudinal direction, and the transversal direction is defined to be transverse to the longitudinal direction, as shown in Fig.~\ref{fig:direction}.
In the literature, longitudinal and rotational trapping with optical radiation pressure is well studied and experimentally tested.
For longitudinal trapping, the optical spring effect is often utilized~\cite{Corbitt:2006zsn, Rehbein:2008yv}.
Radiation pressure in a Fabry-P\'erot cavity behaves similar to a spring at a detuned point.
In the case of rotational degrees of freedom, it is known that the rotational motion of a suspended Fabry-P\'erot cavity is unstable due to the radiation pressure inside the cavity.
This is called Sidles-Sigg instability~\cite{Sidles:2006vzf}.
We can avoid Sidles-Sigg instability by using a triangular cavity, and several experiments have already utilized triangular cavities to stabilize the suspended mirrors~\cite{Matsumoto:2014, Matsumoto:2018via, Komori:2019zlg}.
Another method of mitigating Sidles-Sigg instability is by the angular feedback control with optical radiation pressure~\cite{Enomoto:2016lee, Nagano:2016wcg}.

However, transversal trapping of a disk mirror has not been fully examined experimentally.
Although some configurations were proposed for a stable optical levitation~\cite{Singh2010, guccione, Michimura:2017zfe}, they have not been experimentally realized.
In the configuration proposed in Ref.~\cite{Singh2010}, two horizontal laser beams are introduced as optical tweezers to trap the levitated mirror in the horizontal direction.
In Ref.~\cite{guccione}, the authors proposed to build three Fabry-P\'erot cavities in a tripod-like arrangement below the levitated mirror.
Since the Fabry-P\'erot cavities are slightly inclined from the vertical axis, the optical springs can trap the horizontal motion of the levitated mirror.
In Ref.~\cite{Michimura:2017zfe}, the levitated mirror is sandwiched by two Fabry-P\'erot cavities aligned vertically.
The horizontal motion of the levitated mirror is trapped using the geometry of the Fabry-P\'erot cavities.



In this paper, we report an experimental demonstration of the optical transversal trapping of a mirror.
Our system utilizes a torsional pendulum as a force sensor to measure the optical trapping force acting on the mirror.
Here, we consider the sandwich configuration~\cite{Michimura:2017zfe} because it employs a transversal restoring force due to the optical radiation pressure in the Fabry-P\'erot cavities aligned vertically, which is explained in further detail in the next section.
We conducted measurements to confirm whether the horizontal restoring force on the mirror in a Fabry-P\'erot cavity is significantly positive.





\section{\label{sec:level1}Sandwich configuration}
\begin{figure}
    \centering
        \includegraphics[width=\columnwidth,clip]{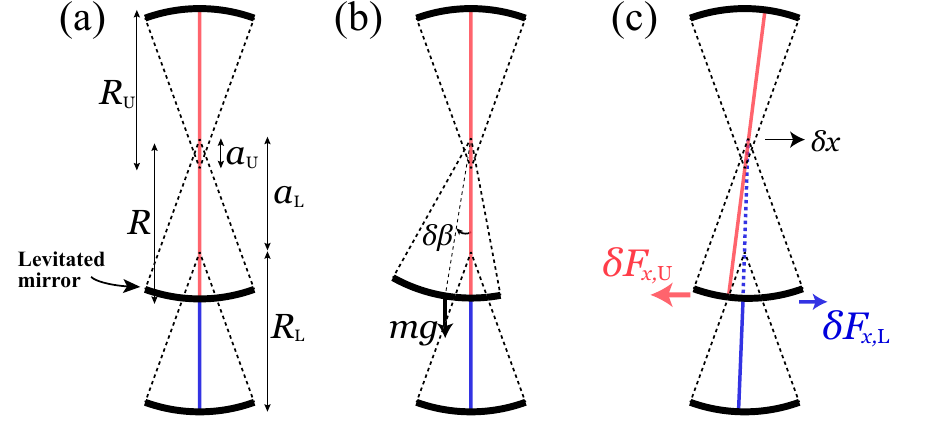}
        \caption{Schematic of the sandwich configuration and its stability.
        A levitated mirror is sandwiched by the two cavities.
        The radiation pressure of the lower cavity supports the levitated mirror, whereas the upper cavity stabilizes the levitated mirror.
        $a_\mathrm{U(L)}$ is the distance between the center of curvatures of the upper (lower) cavity. $R_\mathrm{U(L)}$ is the radius of the curvature of the fixed mirror of the upper (lower) cavity. $R$ is the radius of the curvature of the levitated mirror.
        (a) Balanced state. The red (light gray) line indicates the laser beam in the upper cavity, while the blue (dark gray) line indicates the laser beam in the lower cavity.
        Each mirror has a curvature, and the dotted lines show the radii of the curvatures.
        (b) Rotational motion of the levitated mirror. The gravity produces the restoring torque on the levitated mirror.
        (c) Transversal motion of the levitated mirror. The inclined laser beam gives the restoring force on the levitated mirror.
        }
        \label{fig:config}
\end{figure}

The sandwich configuration is a levitation configuration proposed in Ref.~\cite{Michimura:2017zfe}.
Two Fabry-P\'erot cavities are built above and below a levitated mirror as shown in Fig.~\ref{fig:config}.
The bottom face of the levitated mirror is the cavity mirror for both the upper and lower cavities, while the top face of the levitated mirror is anti-reflection coated.
The lower cavity produces radiation pressure that supports the levitated mirror, while the upper cavity is introduced to stabilize the levitated mirror.
The sandwich configuration is a configuration through which the vertical displacement sensitivity of the levitated mirror can reach the standard quantum limit.

To realize levitation, the levitated mirror must be stable in all degrees of freedom, which are $(x, y, z)$ for the position and $(\alpha, \beta, \gamma)$ for the rotation, where $\alpha$, $\beta$, and $\gamma$ are the rotation angles around $x$, $y$, and $z$ axes, respectively.
Here, we set the origin of the coordinates to be the center of the curvature of the levitated mirror instead of the center of mass to diagonalize the equation of motion.
Moreover, the $z$ axis is set to be vertical.
Since the sandwich configuration is symmetric in the $z$ axis, it is sufficient to consider the motion in $(x, z, \beta)$ for the stability.

For stable levitation, all restoring force in each degrees of freedom must be positive.
Let $(\delta F_x, \delta F_z, \delta N_\beta)$ represent the restoring forces and the restoring torque corresponding to the small displacement $(\delta x, \delta z, \delta \beta)$.
The linear response matrix, $K$, is given in the diagonalized form by
\begin{align}
    K = \left(
              \begin{array}{ccc}
                  K^\mathrm{hor}_\mathrm{L} + K^\mathrm{hor}_\mathrm{U} & 0 & 0 \\
                  0 & K^\mathrm{opt}_\mathrm{L} + K^\mathrm{opt}_\mathrm{U} & 0 \\
                  0 & 0 & mgR
              \end{array}
        \right),
    \label{eq:K}
\end{align}
where $(\delta F_x, \delta F_z, \delta N_\beta) = -K(\delta x, \delta z, \delta \beta)$.
Therefore, we can consider the stability of each degree of freedom separately.
There is no optical torque around the center of curvature because the optical paths pass through the center of curvatures of the cavity mirrors.

In the vertical direction, $z$, the levitated mirror can be trapped by double optical springs~\cite{Corbitt:2006zsn}.
$K^\mathrm{opt}_\mathrm{L}$ and $K^\mathrm{opt}_\mathrm{U}$ in Eq.~(\ref{eq:K}) represent the spring constants of the optical springs of the lower and upper cavities, respectively.

In the $\beta$ rotational direction, the levitated mirror is trapped with gravitational potential, similar to an ordinary suspended pendulum, when the mirror is convex downwards as shown in Fig.~\ref{fig:config}.
Accordingly, the spring constant can be expressed as $mgR$, where $m$ is the mass of the levitated mirror, $g$ is the gravitational acceleration, and $R$ is the radius of curvature of the levitated mirror.

The stability in the horizontal direction, $x$, is nontrivial as it is due to the geometrical configuration.
When the levitated mirror moves slightly in the horizontal direction, the optical paths in the cavities incline.
The inclined radiation pressure of the cavities provides restoring force in the horizontal direction.
In the Fourier domain, the spring constant for horizontal direction, $K^\mathrm{hor}_J$, is expressed as
\begin{align}
    K^\mathrm{hor}_J &= k^\mathrm{hor}_J + i\omega \gamma^\mathrm{hor}_J \nonumber \\
    &= \pm \frac{1}{a_J}\frac{2P_J}{c}\left[1 - i\omega \frac{\pi l_J}{\mathcal{F}_J c(1 - G_J)}\right],
    \label{eq:Khor}
\end{align}
where $a_J$ is the distance between the center of curvatures, $P_J$ is the intracavity power, $\mathcal{F}_J$ is the cavity finesse, $l_J$ is the cavity length, and $c$ is the speed of light.
$J = \mathrm{U, L}$ indicates the upper or lower cavity.
$G_J$ is defined as $G_J = (1 - l_J/R_J)(1 - l_J/R)$.
$R_J$ is the radius of curvature of the fixed mirror.
The sign of Eq.~(\ref{eq:Khor}) is positive for the upper cavity and negative for the lower cavity.
The damping term in Eq.~(\ref{eq:Khor}) is derived from the phase delay of the laser light inside the cavity due to the finiteness of the speed of light.
We focus on $k^\mathrm{hor}_J$ because the damping term, $\gamma^\mathrm{hor}_J$, is negligible with realistic parameters aimed at reaching the standard quantum limit~\cite{Michimura:2017zfe}.
The intracavity power in the lower cavity must be larger than that in the upper cavity to support the levitated mirror.
Nevertheless, the total restoring force can be positive if we adopt a sufficiently small $a_\mathrm{U}$.
This horizontal trapping scheme is unique to the sandwich configuration and has not been demonstrated experimentally in the literature.

\section{\label{sec:level1}Experimental method}
\begin{figure*}
    \centering
        \includegraphics[width=2\columnwidth, clip]{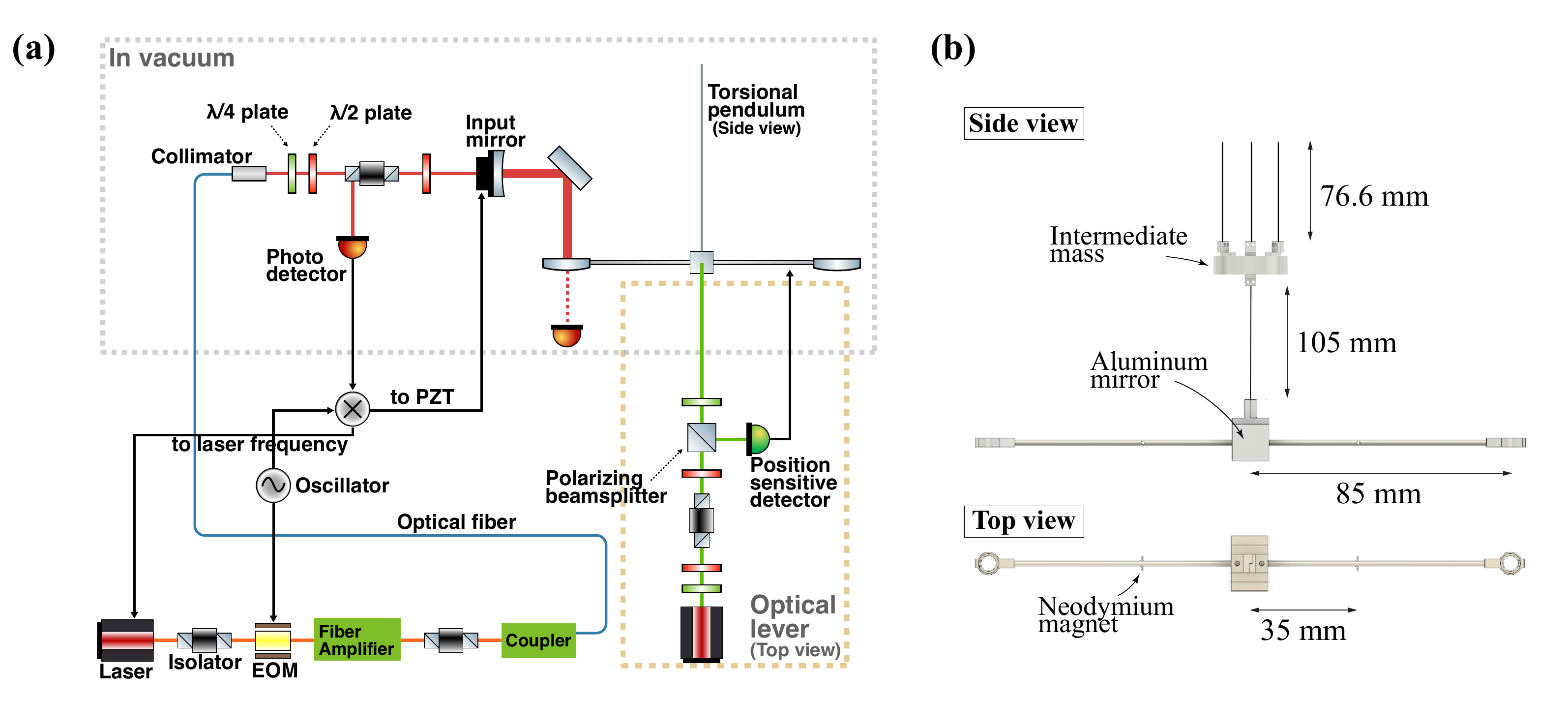}
        \caption{
        (a) Experimental setup.
        The cavity is built on the edge of the torsional pendulum.
        The horizontal rotation of the torsional pendulum is read out with an optical lever.
        The torsional pendulum and the cavity are in the vacuum chamber, and the laser beam is introduced through an optical fiber. EOM, electro-optic modulator; PZT, piezoelectric transducer.
        (b) The torsional pendulum.
        The quarter-inch mirrors are put on the edges of the bar.
        The aluminum mirror attached to the center of the bar is used for the optical lever.
        The bar is suspended with a 105~mm wire.
        The pendulum is a double pendulum to isolate the system from seismic motion, and the intermediate mass is suspended with 76.6~mm wires.
        Two neodymium magnets are attached on both the arms and are 35~mm apart from the center of the bar for coil-magnet actuators.
        }
        \label{fig:setup}
\end{figure*}

The purpose of this experiment is to demonstrate the trapping of a mirror in the horizontal direction with the sandwich configuration.
The trapping is possible if the sandwich configuration provides a positive restoring force to the levitated mirror.
Therefore, we need to evaluate the restoring force in horizontal direction precisely.

To detect the restoring force due to the sandwich configuration, we utilized a torsional pendulum as a force sensor, as shown in Fig.~\ref{fig:setup}.
A mirror was attached to the edge of the torsional pendulum; we treated this mirror as the levitated mirror in the sandwich configuration and built a cavity with this mirror.
Restoring force on the mirror was reflected in the restoring torque of the torsional pendulum.
We note that the rotational motion of the pendulum and the horizontal motion of the mirror are identical for the mirror if the displacement of the motion is small compared to the length of the pendulum bar.
Since a torsional pendulum generally has a small restoring force, it is sensitive to an additional restoring force, which originates from the optical trapping force in the sandwich configuration.
Furthermore, we can omit the lower cavity of the sandwich configuration in our experiment because the restoring force in the horizontal direction is given by the upper cavity, and the torsional pendulum supports the mirror against gravitational force instead of the lower cavity.

The additional restoring force on the pendulum was reflected in the shift in the resonant frequency of the pendulum.
The spring constant 
corresponding to the extra force can be expressed as
\begin{align}
    k_\mathrm{ext} = \frac{(2\pi)^2 I}{L^2}(f_\mathrm{eff}^2-f_0^2),
    \label{eq:optrestoring}
\end{align}
where $I$ is the moment of inertia of the pendulum, $f_0$ is the original resonant frequency of the pendulum, $f_\mathrm{eff}$ is the resonant frequency of the pendulum when the extra force is 
applied, and $L$ is the distance between 
the suspension point of the pendulum bar and the beam spot position on the mirror.
We derived the restoring force due to the sandwich configuration by measuring the shift in the resonant frequency according to Eq.~(\ref{eq:optrestoring}).

The resonant frequency of the pendulum was determined by measuring the 
open-loop transfer functions of the pendulum control. The torsional motion of the pendulum was monitored with an optical lever. The optical lever beam at the wavelength of 850~nm was injected into the center of the pendulum, where an aluminum mirror was attached, and the position of the reflected beam was read out with a position sensitive detector. The torsional mode of the pendulum was feedback controlled by applying differential signals to coil-magnet actuators attached to both arms of the pendulum.
\begin{figure}
    \centering
        \includegraphics[width=\columnwidth,clip]{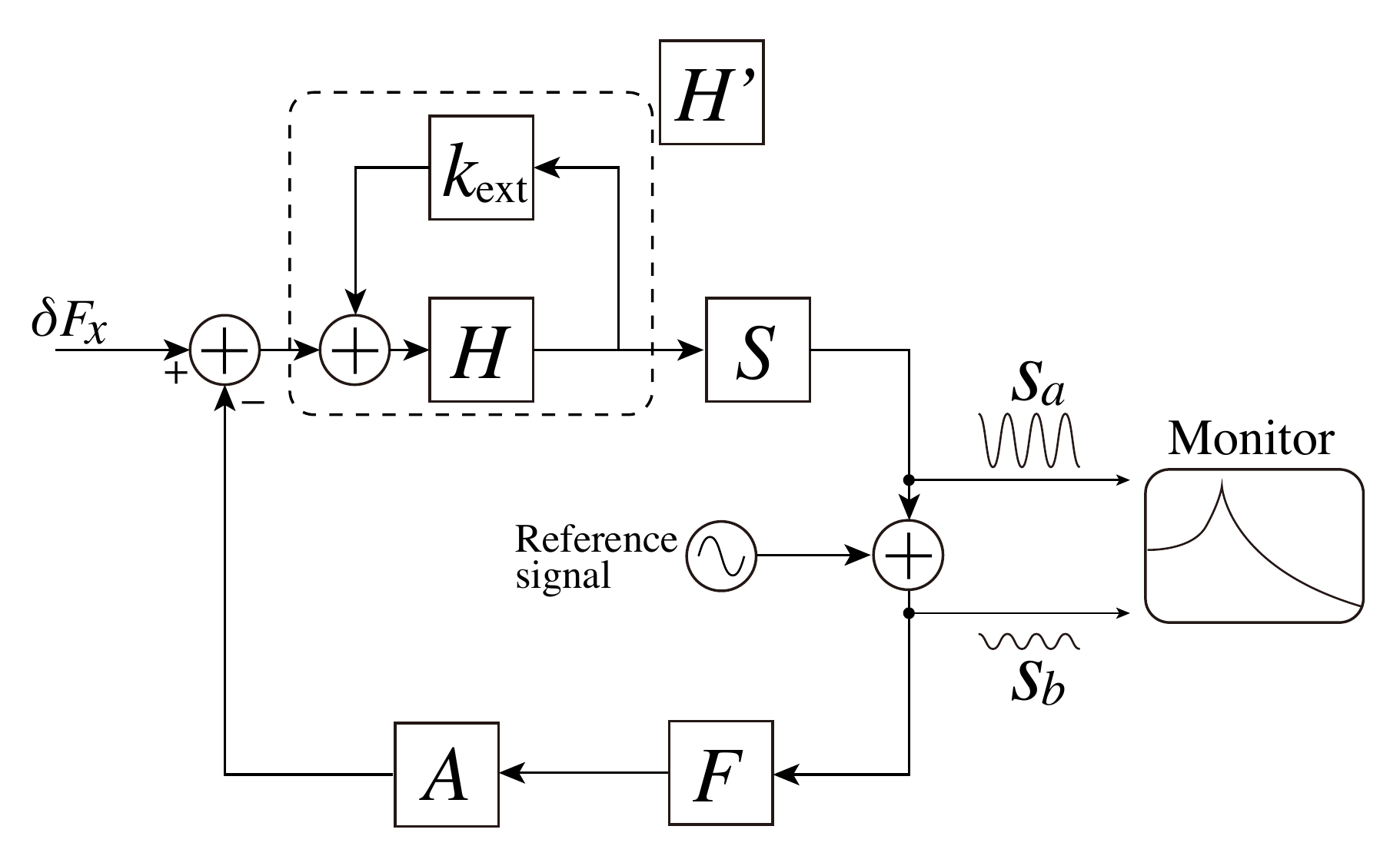}
        \caption{Block diagram of the feedback control.
        When the cavity is not illuminated by the laser beam, there is no $k_\mathrm{ext}$ component. When the cavity is illuminated, the additional restoring force, $k_\mathrm{ext}$, changes the transfer function of the torsional pendulum into $H'$.
        In open-loop transfer function measurement, the reference signal was injected into the feedback control loop.
        $\delta F_x$ is the external force exerting on the torsional pendulum. This external force was suppressed by the feedback control.
        }
        \label{fig:blockdiagram}
\end{figure}
This feedback control is described by the block diagram in Fig.~\ref{fig:blockdiagram}.
$H$, $S$, $F$, and $A$ represent the transfer functions of the torsional pendulum, optical lever sensor, filter circuit, and the actuator, respectively.
$H$ can be expressed as
\begin{align}
    H &= \frac{L^2}{I}\frac{1}{- \omega^2 + \omega_0^2 + i \omega \omega_0/Q},
\end{align}
where $Q$ is the quality factor of the torsional pendulum and $\omega_0 = 2\pi f_0$.
$F$ can be expressed as
\begin{align}
    F &= C \frac{1 + i\omega /(2\pi\times 47.6~\mathrm{mHz})}{[1 + i\omega /(2\pi\times 3.39~\mathrm{Hz})][1 + i\omega /(2\pi\times 4.82~\mathrm{Hz})]},
\end{align}
where $C$ is the gain factor of the filter circuit.
$S$ and $A$ have constant values.
When the cavity is illuminated by the laser beam, an additional restoring force, $k_\mathrm{ext}$, is applied to the pendulum.
As a result, the transfer function of the pendulum transforms into $H'$.
An expression for $H'$ is obtained by replacing $\omega_0$ in $H$ with $\omega_\mathrm{eff} (= 2\pi f_\mathrm{eff}$).
The open-loop transfer functions, $G = HSFA$ and $G' = H'SFA$, are measured.
To measure the open-loop transfer function, a sufficiently large reference signal was injected into the feedback control loop.
The open-loop transfer function can be estimated by taking the ratio, $s_a/s_b$, where $s_a$ and $s_b$ are the back and forth signals of the injected signal, respectively, as shown in Fig.~\ref{fig:blockdiagram}.
With the feedback control, the horizontal motion of the mirror was stabilized to 1.3~$\mu \mathrm{m}$ in the root mean square value, which is enough to form a cavity.

The pendulum was designed to have a small restoring force for it to be a sensitive force sensor.
The moment of inertia of the pendulum was $I = 7.2\times 10^{-6}~\mathrm{kg~m^2}$ and the mass of the pendulum was $8.8~\mathrm{g}$.
A small moment of inertia is preferred for the larger susceptibility of the pendulum.
Therefore, we used a quarter-inch mirror on the edge of the pendulum to reduce the moment of inertia and used a thin aluminum pole for the arm.
The length of the pendulum bar was $2L=17~\mathrm{cm}$.
The suspension wire length is $105~\mathrm{mm}$ and the radius of the wire is $20~\mathrm{\mu m}$.
To minimize the restoring force of the pendulum, we selected an ultra-thin tungsten wire.
Thanks to its tensile strength, this thin wire could suspend the pendulum while keeping the restoring force sufficiently small.

The cavity was illuminated by a laser beam of wavelength 1550~nm.
The maximum output power of the laser source was 2~W.
The radii of the curvature of the pendulum mirror and the input mirror were 75~mm. The distance between the center of curvatures was measured to be $a_\mathrm{U} = 8.9 \pm 0.8 ~\mathrm{mm}$ from transverse mode spacing measurements.
The finesse of the cavity was measured to be $880 \pm 90$.
The cavity length was stabilized to the resonance using a piezoelectric actuator attached to the input mirror. We used the Pound-Drever-Hall method~\cite{Drever:1983las} to obtain the error signal for the cavity length control. We injected different input powers to the cavity to see the power dependence of the trapping force.

The setup needs to be isolated from the environmental disturbances to keep the cavity on the resonance stable and for precise measurement of the transfer function.
The setup was in a vacuum with $1~\mathrm{Pa}$, to prevent air turbulence.
The pendulum is a double pendulum; the intermediate mass is suspended with three wires and has strong eddy-current damping introduced by surrounding neodymium magnets to minimize the translational fluctuation.

\section{\label{sec:level1}Result and discussion}
\begin{figure}
    \centering
        \includegraphics[width=\columnwidth,clip]{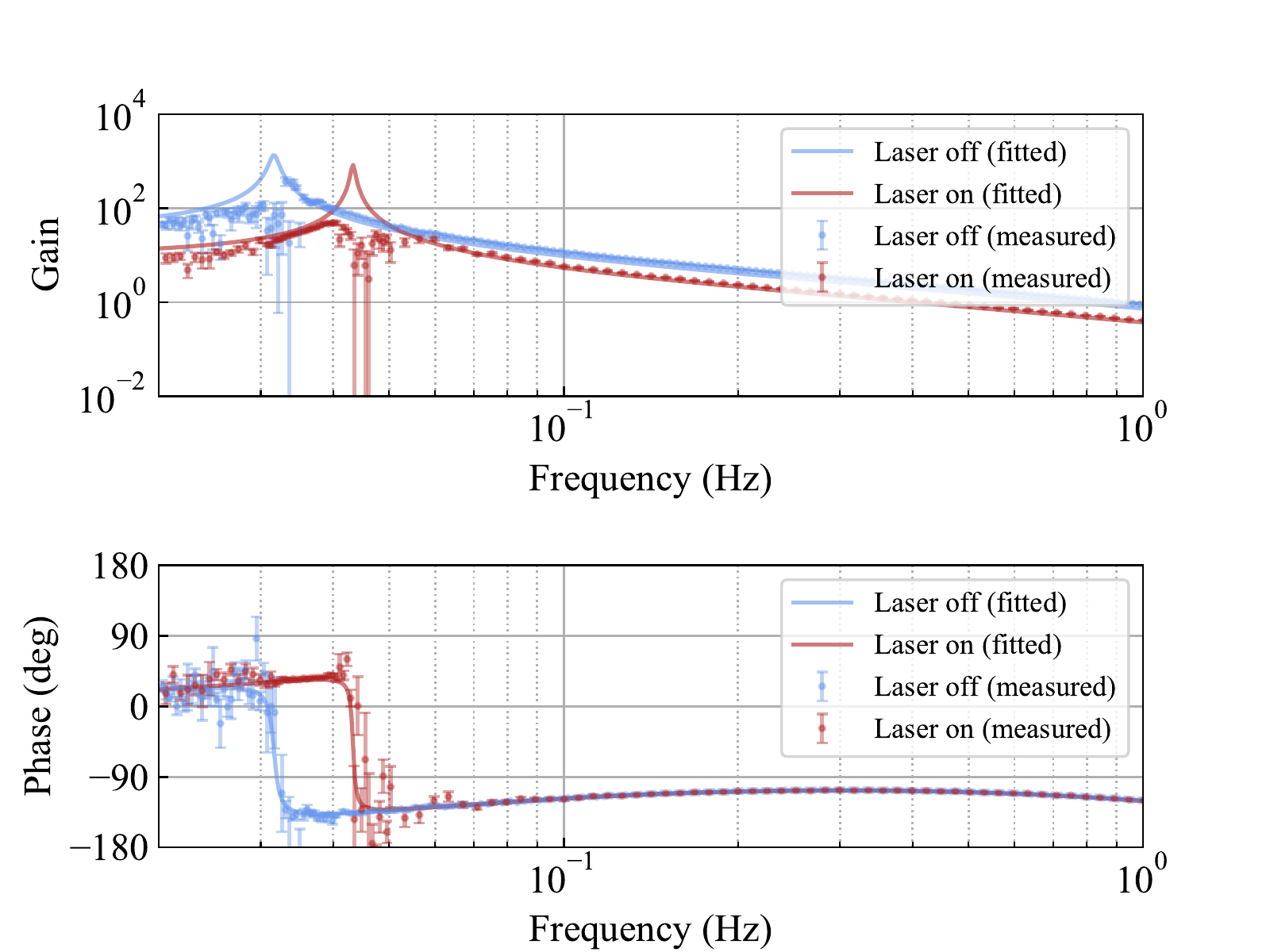}
        \caption{Measured transfer functions of the torsional pendulum.
        The points represent each measured value and the lines represent the fitting.
        The blue (light gray) points and lines represent the data when the laser was off.
        The red (dark gray) points and lines represent the data when the laser was on, with the intracavity power of $29.7 \pm 8.0~\mathrm{W}$.
        The fitted parameters are the resonant frequency, $Q$ value of the resonance, and the overall gain factor.
        The resonant frequencies and the $Q$ values are fitted with the phase data of the measured transfer function.
        The overall gain factors are fitted with the gain data of the measured transfer function.}
        \label{fig:transfunc}
\end{figure}
\begin{figure}
    \centering
        \includegraphics[width=\columnwidth,clip]{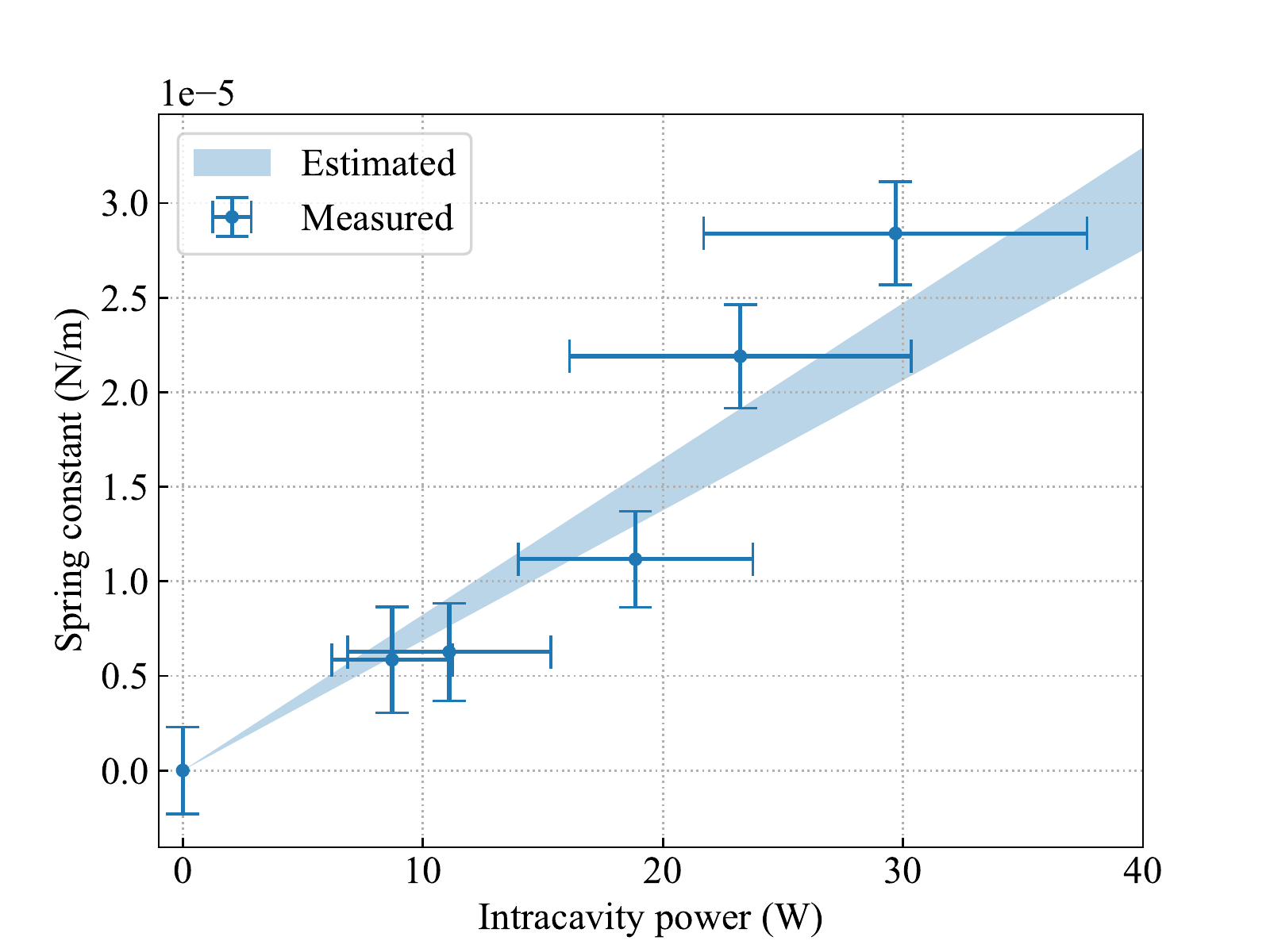}
        \caption{Spring constant corresponding to the extra restoring force, derived from the sandwich configuration.
        First, we measured the restoring force of the torsional pendulum alone.
        The data point with zero intracavity power corresponds to this measurement.
        Then, we injected laser beams with different powers.
        The other five points represent the results with the laser on.
        The shaded region represents the estimated values that are calculated with the cavity parameters based on Eq.~(\ref{eq:Khor}).}
        \label{fig:springconstants}
\end{figure}

We measured the transfer function of the pendulum, which is shown in Fig.~\ref{fig:transfunc}.
Around the resonant frequency, the phase of the transfer function flips, so it can be determined.
The result shows that the resonant frequency of the pendulum alone was $32.2 \pm 1.1~\mathrm{mHz}$.
The resonant frequency shifted to $43.5 \pm 0.4~\mathrm{mHz}$ when the cavity was illuminated by the circulating light with the power of $29.7 \pm 8.0~\mathrm{W}$.
The shift in the resonant frequency indicated an additional positive restoring force on the pendulum. We could estimate the spring constant due to the optical radiation pressure to be $(2.84 \pm 0.27)\times 10^{-5}~\mathrm{N/m}$, according to Eq.~(\ref{eq:optrestoring}).
We found that it was difficult to precisely measure the gain of the transfer functions around the resonance.
This could be because the value of the gain was extremely high.
When the gain is high, the measured signal that is suppressed by the gain factor tends to be too small to precisely determine its amplitude.
Although this measurement was less than ideal, the absence of a peak in the gain data was immaterial to our results because the resonant frequency could be determined by using the phase data.

We also varied the intracavity power and evaluated the spring constants for each power value.
Figure~\ref{fig:springconstants} shows that every data point is consistent with the estimated values calculated with the measured cavity parameter that is the distance between the centers of curvature, $a_\mathrm{U}$.
From this result, we confirmed the extra restoring force was originated from the radiation pressure of the resonating light.
Furthermore, as this result indicated that the theoretical prediction, Eq.~(\ref{eq:Khor}), was valid, it also suggested that even if we had built the lower cavity, the lower cavity would not have affected the horizontal stability of the suspended mirror because $a_\mathrm{L}$ would be much larger than $a_\mathrm{U}$.

The intracavity power was estimated by measuring the power of the transmitted light of the Fabry-P\'erot cavity with a photo detector.
The transmissivity of the mirror on the torsional pendulum 
was $0.05 \pm 0.01\%$.

\begin{figure}
    \centering
        \includegraphics[width=\columnwidth,clip]{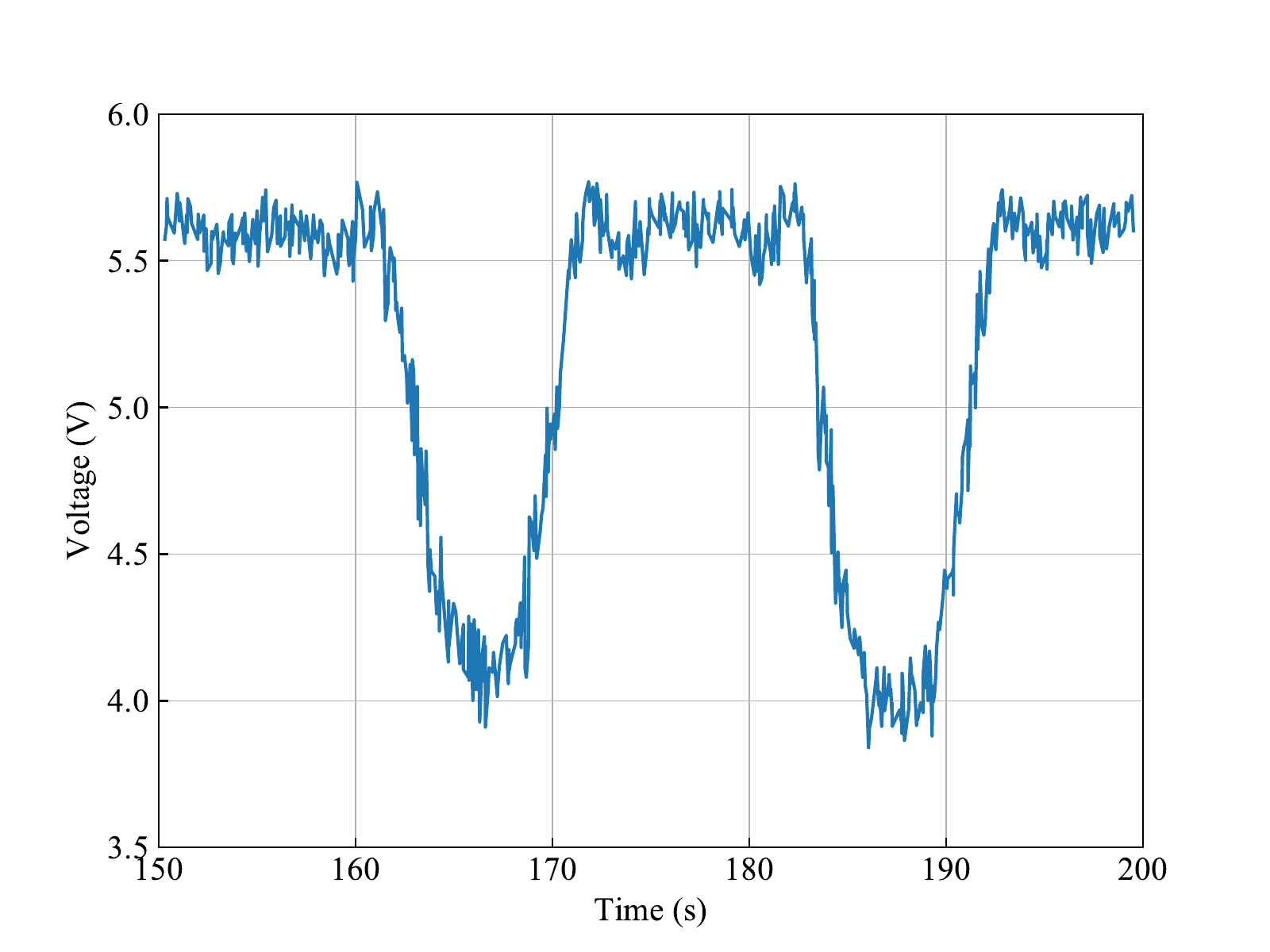}
        \caption{Output of the photo detector monitoring the transmitted light of the cavity, with a intracavity power of $23.2 \pm 7.1~\mathrm{W}$. The power of the transmitted light signifies the intracavity power. The signal fluctuated at a frequency of 50~mHz, i.e., the frequency at which we swung the torsional pendulum to measure the open-loop transfer function.
        }
        \label{fig:transtimeseries}
\end{figure}

The major source of uncertainty in the measured intracavity power comes from the fluctuation in the intracavity power when we actuate the torsional pendulum to measure the transfer functions, as shown in Fig.~\ref{fig:transtimeseries}.
The period of the fluctuation varied corresponding to the actuation period of the torsional pendulum.

We need to swing the pendulum with a larger amplitude than the pendulum movement with no feedback control.
To reduce uncertainty in the intracavity power estimation, it is necessary to reduce the pendulum movement by, for example, creating vibration isolation around the resonant frequency of the pendulum.
We note that the swinging amplitude was identical for all measurements, and thus in principle, the fluctuation in the intracavity power was proportional to its magnitude.

We repeated the measurement of the transfer function three times for each intracavity power.
We obtained the estimated value of the resonant frequency for every measured transfer function.
The statistical uncertainty in the resonant frequency was estimated, and it was found to be the dominant source of uncertainty in estimating the spring constant.

\section{\label{sec:level1}Conclusion}
We proved the stability of the sandwich configuration in the horizontal direction.
To measure the restoring force on the mirror with the sandwich configuration, we utilized a torsional pendulum as a force sensor.
By observing the shift in the resonant frequency of the torsional pendulum, we evaluated the additional restoring force that originates from the sandwich configuration.
The measured restoring force was positive and the dependence on the intracavity power was consistent with the theoretical prediction.

Our work demonstrates the first experimental validation of 
the transversal trapping for an optically levitated mirror of a Fabry-P\'erot cavity.
Consequently, we showed that all degrees of freedom could be trapped stably by optical levitation.
Our study facilitates the future realization of optical levitation in the sandwich configuration.
Our research is a crucial step towards 
realizing and elucidating macroscopic quantum systems.
We have also established a method to measure the horizontal restoring force that acts on a levitated mirror, and this method can also be applied to other levitation configurations for evaluating their stability.

\begin{acknowledgments}
We thank Kentaro Komori, Yutaro Enomoto, Ooi Ching Pin, and Satoru Takano for fruitful discussions.
This work was supported by a Grant-in-Aid for Challenging Research Exploratory Grant No. 18K18763 from the Japan Society for the Promotion of Science (JSPS) and by JST CREST Grant No. JPMJCR1873.
T.K. was supported by KAKENHI Grant No. 19J21861 from the JSPS.
N.K. was supported by KAKENHI Grant No. 20J01928 from the JSPS.
\end{acknowledgments}


\bibliography{main}

\end{document}